\documentclass[aps,pra,twocolumn,amsfonts,amssymb,amsmath,showpacs,
floatfix,nofootinbib,groupedaddress,superscriptaddress,citesort]{revtex4}
\usepackage{mathrsfs}
\usepackage{amsfonts}
\usepackage{amstext}
\usepackage{amsmath}
\usepackage{amssymb}
\usepackage{bm}
\usepackage{CJK}
\usepackage{bbm}
\usepackage[dvips]{graphicx}
\def\qed{\leavevmode\unskip\penalty9999 \hbox{}\nobreak\hfill
     \quad\hbox{\leavevmode  \hbox to.77778em{%
              \hfil\vrule   \vbox to.675em%
               {\hrule width.6em\vfil\hrule}\vrule\hfil}}
     \par\vskip3pt}

\begin{document}

\title{When quantum channel preserves product states}

\thanks{Dedicated to Prof. Jinchuan Hou, on the occasion of his
60th birthday}

\author{Yu Guo}
\email{guoyu3@aliyun.com} \affiliation{School of Mathematics
and Computer Science, Shanxi
Datong University, Datong 037009, China}%

\author{Zhaofang Bai}
\thanks{Corresponding author}
\email{baizhaofang@xmu.edu.cn}\affiliation{School of Mathematical
Sciences, Xiamen University, Xiamen 361000, China}
\author{Shuanping Du}
\email{shuanpingdu@yahoo.com}\affiliation{School of Mathematical
Sciences, Xiamen University, Xiamen 361000,  China}

\author{Xiulan Li}
\affiliation{School of Mathematics and Computer Science, Shanxi Datong University, Datong 037009, China}%

\begin{abstract}

Product states are always considered as the states that don't
contain quantum correlation. We discuss here when a quantum
channel sends the product states to themselves. The exact forms of
such channels are proposed. It is shown that such a quantum
channel is a local quantum channel, a composition of a local
quantum channel and a flip operation, or such that one of the
local states is fixed. Both finite- and infinite-dimensional
systems are considered.

\end{abstract}


\maketitle


Quantum systems can be correlated in ways
inaccessible to classical objects.
This quantum
feature of correlations
not only is the key to our understanding
of quantum world, but also is essential for the
powerful applications of quantum information and quantum
computation.
Product state is the state without any quantum correlation \cite{Horodecki,Guhne}.
It is the only state that has zero mutual information \cite{Jevtic} which is interpreted as a measure of total correlations
between its two subsystems.
It neither contains quantum discord (QD) \cite{Ollivier} nor contains the measurement-induced nonlocality (MIN) \cite{Luo,Guo3}.
Recently, it has been shown that the super discord \cite{Singh} of $\rho_{ab}$
is zero if and only if it is a product state \cite{Libo}.

In particular, it is crucial to study the behavior of quantum correlation under the
influence of noisy channel ~\cite{Streltsov,Filippov,Zyczkowski2,Rao,Cui,Shabani,Altinatas,Mazzola,Ciccarello,
Hu,Guo3,Guo1,Guo2,Hassan}. For example, local channel that cannot create QD is investigated in \cite{Streltsov,Hu,Guo1},
local channel that preserves the state with vanished MIN is characterized in \cite{Guo3} and
local channel that preserves the maximally entangled states is explored in \cite{Guo2}.
The goal of this paper is to discuss when a quantum channel preserves the product states.

We fix some notations first. Let $H$, $K$ be separable complex
Hilbert spaces, and $\mathcal{B}(H,K)$ ($\mathcal{B}(H)$ when $K =
H$) be the Banach space of all (bounded linear) operators from $H$
into $K$. Recall that $A \in \mathcal{B}(H)$ is self-adjoint if $A
= A^\dag$ ($A^\dag$ stands for the adjoint operator of $A$); and
$A$ is positive, denoted by $A\geq 0$, if $A$ is self-adjoint with
the spectrum falling in the interval $[0,\infty)$ (or
equivalently, $\langle\psi|A|\psi\rangle\geq 0$ for all
$|\psi\rangle\in H$). A linear map
$\phi:~\mathcal{B}(H)\rightarrow \mathcal{B}(K)$ is called a
positive map if $A\geq0$ implies $\phi(A)\geq0$ for any
$A\in\mathcal{B}(H)$. Let $\mathcal{M}_n(\mathcal{B}(H))$ be the
algebra of all $n$ by $n$ matrices with entries are operators in
$\mathcal{B}(H)$. Let
$\mathbbm{1}_n\otimes\phi:\mathcal{M}_n(\mathcal{B}(H))\rightarrow
\mathcal{M}_n(\mathcal{B}(K))$ be the map defined by
$(\mathbbm{1}_n\otimes \phi)[X_{ij}]=[\phi(X_{ij})]$. We call that
$\phi$ is completely positive if $\mathbbm{1}_n\otimes \phi$ is
positive for any $n$.

We review the definition of the quantum channel.
Let $\mathcal{T}(H)$, $\mathcal{T}(K)$ be the trace
classes on $H$, $K$ respectively.
Recall that
a quantum channel is described by a trace-preserving completely
positive linear map $\phi:~\mathcal{T}(H)\rightarrow
\mathcal{T}(K)$.
Every quantum channel $\phi$ between two
systems respectively associated with Hilbert spaces $H$ and $K$ admits the form \cite{Hou}
\begin{eqnarray}
\phi(\cdot)=\sum\limits_{i} X_i(\cdot) X_i^\dag, \label{channel}
\end{eqnarray}
where $\{X_i\}\subset \mathcal{B}(H,K)$ satisfies that $\sum_{i}
X_i^\dag X_i=I_H$, $I_H$ is the identity operator on $H$. If $\dim
H=+\infty$ and $\dim K=+\infty$, then there may have infinite
$X_i$s in Eq.~(\ref{channel}). We call $\phi$ is a
\emph{completely contractive channel } if $\phi(\mathcal{S}(H))$
is a single state \cite{Zyczkowski}, i.e. there exists a fixed
state $\omega_0\in\mathcal{S}(H)$ such that
\begin{eqnarray}
\phi(\cdot)={\rm Tr}(\cdot)\omega_0.
\end{eqnarray}

Let $H_{ab}=H_a\otimes H_b$ with $\dim H_a\leq+\infty$ and $\dim
H_b\leq+\infty$ be the state space of the bipartite system A+B.
Let $\mathcal{S}(H_{ab})$ and $\mathcal{S}_P(H_{ab})$ be the set
of all quantum states acting on $H_{ab}$ and the set of all
product states in $\mathcal{S}(H_{ab})$ respectively. That is
$\mathcal{S}_P(H_{ab})=\{\rho\otimes\delta:
\rho\in\mathcal{S}(H_a), \delta\in\mathcal{S}(H_b)\}$. Let
$\{|i\rangle\}$ and $|j'\rangle$ be the orthonormal bases of $H_a$
and $H_b$ respectively. The operator
$F=\sum_{i,j}|i\rangle|j'\rangle\langle j'|\langle i|$ is called
the swap operator from $H_{ba}=H_b\otimes H_a$ to $H_{ab}$
\cite{Guo4}, namely,
$F|\psi_b\rangle|\psi_a\rangle=|\psi_a\rangle|\psi_b\rangle$ for
any $|\psi_b\rangle|\psi_a\rangle\in H_{ba}$ (note that $F$ is an
isometry since $F^\dag F=I_{ba}$). Then $F\delta\otimes\rho
F=\rho\otimes\delta\in\mathcal{S}_P(H_{ab})$ for any
$\delta\otimes\rho\in\mathcal{S}_P(H_{ba})$. We denote by
$f(\delta\otimes\rho):=F\delta\otimes\rho F$.

The following is the main result of this paper.

{\it Theorem 1.} Let $\phi:\mathcal{T}(H_{ab})\rightarrow\mathcal{T}(H_{ab})$ be a quantum channel.
Then $\phi(\mathcal{S}_P(H_{ab}))\subseteq\mathcal{S}_P(H_{ab})$ if and only if it has
one of the following forms.

(i) $\phi=\phi_a\otimes \phi_b$, where $\phi_a$ and $\phi_b$ denote the local quantum channels on part A and B respectively;

(ii) $\phi=f\circ(\psi_a\otimes \psi_b)$, where $\psi_a$ is a quantum channel from $\mathcal{T}(H_a)$ to $\mathcal{T}(H_b)$ and $\psi_b$
is a quantum  channel from $\mathcal{T}(H_b)$ to $\mathcal{T}(H_a)$;

(iii) $\phi(\cdot)=\sigma\otimes\Lambda_b(\cdot)$, where $\sigma$
is a state of part A, $\Lambda_b$ is a quantum channel from
$\mathcal{T}(H_{ab})$ to $\mathcal{T}(H_b)$;

(iv) $\phi(\cdot)=\Lambda_a(\cdot)\otimes\tau$, where $\tau$ is a
state of part B, $\Lambda_a$ is a quantum channel from
$\mathcal{T}(H_{ab})$ to $\mathcal{T}(H_a)$.

Theorem 1 implies that a quantum channel sends product states to
product states if and only if it is a action of two local
operations on part A and part B respectively or is a action of
quantum channel from the total system to a subsystem with another
reduced state fixed. (In Theorem 1, for the notations
$\sigma\otimes\Lambda_b(\cdot)$ and $\Lambda_a(\cdot)\otimes\tau$,
with some abuse of terminology, $\sigma$ can be viewed as a
completely contractive quantum channel from A+B to a single state
$\sigma$ of part A and $\tau$ can be viewed as a completely
contractive quantum channel from A+B to a single state $\tau$ of
part B.)

In order to prove Theorem 1, the following lemmas are necessary.

{\it Lemma 1.} Let
$\Lambda_{a,b}:\mathcal{T}(H_{ab})\rightarrow\mathcal{T}(H_{a,b})$
be a quantum channel. Then $\phi$ as (iii) or (iv) above is a
quantum channel on $\mathcal{T}(H_{ab})$.

{\it Proof.}\quad We check the case of (iii), the case of (iv) can
be argued similarly. We only need to show
$\mathbbm{1}_n\otimes\phi:\mathcal{M}_n(\mathcal{T}(H_{ab}))\rightarrow\mathcal{M}_n(\mathcal{T}(H_{ab}))$
is positive for any $n$. For any positive operator $[S_{ij}]\geq0$
in $\mathcal{M}_n(\mathcal{T}(H_{ab}))$, we have
$(\mathbbm{1}_n\otimes\phi)[S_{ij}]=[\sigma\otimes\Lambda_b(S_{ij})]$.
Note that $[\sigma\otimes\Lambda_b(S_{ij})]\geq0$ if and only if
$\sigma\otimes[\Lambda_b(S_{ij})]\geq0$. Therefore
$(\mathbbm{1}_n\otimes\phi)[S_{ij}]$ is positive since
$[\Lambda_b(S_{ij})]\geq0$. The proof is
completed.\hfill$\blacksquare$

{\it Lemma 2.} Let $K_{a,b}$ be separable complex Hilbert space and let
${\phi}_{a,b}:\mathcal{T}(H_{a,b})\rightarrow\mathcal{T}(K_{a,b})$
be a trace-preserving positive map. If $\phi_a\otimes \phi_b$ is a quantum channel from
$\mathcal{T}(H_{ab})$ to $\mathcal{T}(K_{ab})$, then
$\phi_{a,b}$ is a quantum channel from $H_{a,b}$ to $K_{a,b}$.

{\it Proof.}\quad Let $[S_{ij}]\in\mathcal{M}_n(\mathcal{T}(H_a))$, $[S_{ij}]\geq0$.
Then $[S_{ij}\otimes\rho_0]\geq0$ for any $\rho_0\in\mathcal{S}(H_b)$.
Thus $[\phi_a(S_{ij})\otimes\phi_b(\rho_0)]\geq0$.
For any $|x_0\rangle$, $|y_0\rangle\in K_a$, one has
\begin{eqnarray*}
&&(\mathbbm{1}_n\otimes |x_0\rangle\langle y_0|\otimes \mathbbm{1}_{K_b})[\phi_a(S_{ij})\otimes\phi_b(\rho_0)]\\
&&\cdot(\mathbbm{1}_n\otimes |y_0\rangle\langle x_0|\otimes \mathbbm{1}_{K_b})\\
&=&[\phi_a(S_{ij})\otimes\langle y_0|\phi_b(\rho_0)|y_0\rangle|x_0\rangle\langle x_0|]\\
&=&\langle y_0|\phi_b(\rho_0)|y_0\rangle[\phi_a(S_{ij})\otimes|x_0\rangle\langle x_0|]\geq0,
\end{eqnarray*}
where $\mathbbm{1}_{K_b}$ denotes the identity map on $\mathcal{B}(K_b)$.
Hence $[\phi_a(S_{ij})]\geq0$, that is $\phi_a$ is completely positive, thus it is a quantum channel.
Using similar argument, we can obtain $\phi_b$ is also a quantum channel.
\hfill$\blacksquare$

We are now ready for the proof of Theorem 1.

{\it Proof of Theorem 1.}\quad
By Lemma 1, the `if' part is obvious. We check the `only if' part below.
$\phi(\mathcal{S}_P(H_{ab}))$ has at most three different cases:
(1) There exists a state $\sigma\in\mathcal{S}(H_a)$ such that
$\phi(\mathcal{S}_P(H_{ab}))\subseteq\sigma\otimes \mathcal{S}(H_b)$;
(2) There exists a state $\tau\in\mathcal{S}(H_b)$ such that
$\phi(\mathcal{S}_P(H_{ab}))\subseteq \mathcal{S}(H_a)\otimes\tau$;
(3) There exist $\rho_1\otimes\delta_1$ and $\rho_2\otimes\delta_2$
such that $\phi(\rho_i\otimes\delta_i)=\sigma_i\otimes \tau_i$, $i=1$, 2,
with
$\sigma_1$ and $\sigma_2$ are linearly independent and
$\tau_1$ and $\tau_2$ are linearly independent.

\emph{Case 1.} Since for any
$\gamma\otimes\omega\in\mathcal{S}_P(H_{ab})$,
$\phi(\gamma\otimes\omega)=\sigma\otimes\pi_{\gamma\otimes\omega}$
we let $\Lambda_b(\gamma\otimes\omega)=\pi_{\gamma\otimes\omega}$.
Then $\Lambda_b={\rm Tr}_a\circ\phi$ is is a quantum channel from
$\mathcal{T}(H_{ab})$ to $\mathcal{T}(H_b)$, where ${\rm Tr}_a$
denotes the partial reduction map up to part A, i.e., $\phi$ has
the form in item (iii).

\emph{Case 2.} Similar to Case 1, we can get
$\phi=\Lambda_a\otimes \tau$ is a quantum channel on
$\mathcal{T}(H_{ab})$ with $\Lambda_a$ is a quantum channel from
$\mathcal{T}(H_{ab})$ to $\mathcal{T}(H_a)$, which is the form in
item (iv).

\emph{Case 3.} Let $\phi(\rho_1\otimes\delta_2)=\sigma_3\otimes\tau_3$.
Then either (3.1) $\sigma_1$ and $\sigma_3$ are linearly dependent or (3.2) $\tau_1$ and $\tau_3$
are linearly dependent.

\emph{Case 3.1.} If $\sigma_1$ and $\sigma_3$ are linearly dependent, then $\sigma_1=\sigma_3$ and $\tau_3=\tau_2$.
Let
\begin{eqnarray*}
\mathcal{L}_\rho=\{\rho\otimes\delta:\delta\in\mathcal{S}(H_b)\},\\
\mathcal{R}_\delta=\{\rho\otimes\delta:\rho\in\mathcal{S}(H_a)\}.
\end{eqnarray*}
Next we show that for any $\rho\in{\mathcal{S}(H_a)}$, $\phi(\mathcal{L}_\rho)\subseteq\mathcal{L}_\sigma$ for some $\sigma$ (depending on $\rho$),
and that for any $\delta\in\mathcal{S}(H_b)$, $\phi(\mathcal{R}_\delta)\subseteq\mathcal{R}_\tau$ for some $\tau$  (depending on $\delta$).

Let $\phi(\rho_2\otimes\delta_1)=\sigma_4\otimes\tau_4$,
then either $\sigma_2$ and $\sigma_4$ are linearly dependent or
$\tau_2$ and $\tau_4$ are linearly dependent.
If $\sigma_4\neq\sigma_2$, then $\tau_4=\tau_2$ and thus
$\frac{1}{4}\phi((\rho_1+\rho_2)\otimes (\delta_1+\delta_2))
=\frac{1}{4}(\sigma_1\otimes (\tau_1+\tau_2)+\sigma_2\otimes\tau_2+\sigma_4\otimes \tau_2)$
is not a product state,
which implies that $\sigma_4=\sigma_2$.
Thus
\begin{eqnarray*}
&&\frac{1}{4}\phi((\rho_1+\rho_2)\otimes (\delta_1+\delta_2))\\
&=&\frac{1}{4}(\sigma_1\otimes(\tau_1+\tau_2)+\sigma_2\otimes(\sigma_4+\sigma_2))
\end{eqnarray*}
is a product state
leads to $\tau_4=\tau_1$.
That is, $\phi(\rho_2\otimes\delta_1)=\sigma_2\otimes\tau_1$.
Assume that $\phi(\rho_1\otimes\delta)=\eta\otimes\xi$, $\delta\in\mathcal{S}(H_a)$,
then
(a) either $\sigma_1$ and $\eta$ are linearly dependent or $\tau_1$ and $\xi$ are linearly dependent,
and (b)
either $\sigma_1$ and $\eta$ are linearly dependent or $\tau_2$ and $\xi$ are linearly dependent.
It turns out that
$\eta=\sigma_1$.
That is
\begin{eqnarray*}
\phi(\mathcal{L}_{\rho_1})\subseteq\mathcal{L}_{\sigma_1}.
\end{eqnarray*}
Similarly, we have
\begin{eqnarray*}
\phi(\mathcal{L}_{\rho_2})\subseteq\mathcal{L}_{\sigma_2},\
\phi(\mathcal{R}_{\delta_1})\subseteq\mathcal{R}_{\tau_1},\
\phi(\mathcal{R}_{\delta_2})\subseteq\mathcal{R}_{\tau_2}.
\end{eqnarray*}
For arbitrarily given $\rho\in\mathcal{S}(H_a)$,
let
\begin{eqnarray*}
\phi(\rho\otimes\delta_1)=\xi\otimes \tau_1,\ \phi(\rho\otimes\delta_2)=\varsigma\otimes\tau_2.
\end{eqnarray*}
Then $\xi=\varsigma$.
For any $\delta\in\mathcal{S}(H_b)$, let
\begin{eqnarray*}
\phi(\rho\otimes\delta)=\sigma\otimes \tau.
\end{eqnarray*}
Then
either $\xi$ and $\sigma$ are linearly dependent or $\tau$ and $\tau_1$ are linearly dependent,
and either $\xi$ and $\sigma$ are linearly dependent or $\tau$ and $\tau_2$ are linearly dependent.
We thus can conclude that $\xi=\sigma$.
That is, for any $\rho\in\mathcal{S}(H_a)$,
\begin{eqnarray}
\phi(\mathcal{L}_\rho)\subseteq\mathcal{L}_\sigma \ \mbox{\rm for some}\ \sigma\in\mathcal{S}(H_a).\label{1}
\end{eqnarray}
Similarly, for any $\delta\in\mathcal{S}(H_b)$,
\begin{eqnarray}
\phi(\mathcal{R}_\delta)\subseteq\mathcal{R}_\tau \ \mbox{\rm for some}\ \tau\in\mathcal{S}(H_b).\label{2}
\end{eqnarray}

From Eqs.~(\ref{1}) and (\ref{2}), we can let
\begin{eqnarray}
\phi_a: \rho\mapsto\sigma
\end{eqnarray}
and
\begin{eqnarray}
\phi_b: \delta\mapsto\tau.
\end{eqnarray}
It is clear that $\phi=\rho_a\otimes\phi_b$.
By lemma 2, $\phi_a$ and $\phi_b$ are quantum channels
on $\mathcal{T}(H_a)$ and $\mathcal{T}({H_b})$ respectively.

\emph{Case 3.2.} If $\tau_1$ and $\tau_3$
are linearly dependent, using the similar argument as Case 3.1, we can conclude that
for any $\rho\in\mathcal{S}(H_a)$,
\begin{eqnarray}
\phi(\mathcal{L}_\rho)\subseteq\mathcal{R}_\tau \ \mbox{\rm for some}\ \tau\in\mathcal{S}(H_b).\label{x}
\end{eqnarray}
Similarly, for any $\delta\in\mathcal{S}(H_b)$,
\begin{eqnarray}
\phi(\mathcal{R}_\delta)\subseteq\mathcal{L}_\sigma \ \mbox{\rm for some}\ \sigma\in\mathcal{S}(H_a).\label{y}
\end{eqnarray}
We thus can let
\begin{eqnarray}
\psi_b: \rho\mapsto\tau
\end{eqnarray}
and
\begin{eqnarray}
\psi_a: \delta\mapsto\sigma.
\end{eqnarray}
Therefore $\phi$ has the desired form as item (ii).
We now complete the proof.
\hfill$\blacksquare$

Furthermore, if the output states of a channel $\phi$ on
$\mathcal{T}(H_{ab})$ are always product states, i.e.,
$\phi(\mathcal{S}(H_{ab}))\subseteq\mathcal{S}_P(H_{ab})$, one can
easily conclude that $\phi$ has the form as item (iii) or (iv) in
Theorem 1.

{\it Proposition 1.} Let $\phi:\mathcal{T}(H_{ab})\rightarrow\mathcal{T}(H_{ab})$ be a quantum channel.
Then $\phi(\mathcal{S}(H_{ab}))\subseteq\mathcal{S}_P(H_{ab})$ if and only if it admits the form as item (iii) or (iv) in Theorem 1.

In summary, the quantum channel that preserves product states is characterized mathematically.
It is in nature a combination of two local quantum channels or a
quantum channel such that one of the local states is fixed. Moreover,
if the latter occurs, then it sends any state to a product state and vice versa.
Especially, if $\phi_a$ ($\phi_b$) or $\psi_a$ ($\psi_b$) in Theorem 1
is completely contractive, then $\phi$ sends any state to product state as well.

\smallskip

This work is partially supported by the Natural Science Foundation
of China (Grant No. 11301312, Grant No. 11001230, Grant No.
11171249), the Natural Science Foundation of Shanxi (Grant No.
2013021001-1,  Grant No. 2012011001-2), the Natural Science
Foundation of Fujian (2013J01022, 2014J01024) and the Research
start-up fund for Doctors of Shanxi Datong University (Grant No.
2011-B-01).



\end{document}